\documentclass[11pt]{article}

\usepackage[utf8]{inputenc}
\usepackage[T1]{fontenc}
\usepackage{graphicx}
\usepackage{hyperref}
\usepackage{amsmath, amssymb}
\usepackage[numbers]{natbib}
\usepackage{listings}
\title{Human Oversight-by-Design for Accessible Generative IUIs}

\author{
Blessing Jerry \quad Lourdes Moreno \quad Paloma Mart\'inez\\
\small Computer Science and Engineering Department, Universidad Carlos III de Madrid, Spain\\
\small \texttt{bjerry@pa.uc3m.es, lmoreno@inf.uc3m.es, pmf@inf.uc3m.es}
}

\date{}

\begin{document}
\maketitle

\begin{abstract}
  LLM-generated interfaces are increasingly used in high-consequence workflows (e.g., healthcare communication), where how information is presented can impact downstream actions. These interfaces and their content support human interaction with AI-assisted decision-making and communication processes and should remain accessible and usable for people with disabilities. Accessible plain-language interfaces serve as an enabling infrastructure for meaningful human oversight. In these contexts, ethical and trustworthiness risks, including hallucinations, semantic distortion, bias, and accessibility barriers, can undermine reliability and limit users’ ability to understand, monitor, and intervene in AI-supported processes. Yet, in practice, oversight is often treated as a downstream check, without clear rules for when human intervention is required or who is accountable. We propose oversight-by-design: embedding human judgment across the pipeline as an architectural commitment, implemented via escalation policies and explicit UI controls for risk signalling and intervention. Automated checks flag risk in generated UI communication that supports high-stakes workflows (e.g., readability, semantic fidelity, factual consistency, and standards-based accessibility constraints) and escalate to mandatory Human-in-the-Loop (HITL) review before release when thresholds are violated, or uncertainty is high. Human-on-the-Loop (HOTL) supervision monitors system-level signals over time (alerts, escalation rates, and compliance evidence) to tune policies and detect drift. Structured review feedback is translated into governance actions (rule and prompt updates, threshold calibration, and traceable audit logs), enabling scalable intervention and verifiable oversight for generative UI systems that support high-stakes workflows.
\end{abstract}

\noindent\textit{This is a preprint of a paper accepted for publication in CEUR Workshop Proceedings (IUI Workshops 2026).}

\maketitle

\section{Introduction}

Large language models (LLMs) are rapidly being integrated into systems that operate in high-stakes domains such as healthcare, legal services, and public services. In these contexts, failures are not merely technical errors, but can threaten human safety, autonomy, and rights. For Intelligent User Interfaces (IUIs) that generate interface content and interaction cues, human oversight is not only a governance requirement but an interaction design problem: the system must expose risk, uncertainty, and intervention pathways through UI-level controls  \cite{Parasuraman2000Levels, Shneiderman2020HCAI, Zhang2020ConfidenceExplanation, Kim2024ImNotSureBut}. Accessibility and plain language serve a dual role: they are quality and compliance criteria for generated interface content, and they are prerequisites for enabling diverse reviewers to effectively perform oversight and intervention tasks. Regulatory frameworks such as the EU AI Act require appropriate human oversight for high-risk AI systems to reduce risks to health, safety, and fundamental rights \cite{EU2024AIAct}. However, despite growing consensus on its importance, human oversight remains poorly specified in system design practice, and the Act provides limited guidance on how to implement effective oversight in concrete system architecture \cite{Enqvist2023HumanOversight, HoDacMartinez2024HumanOversight, Hummel2025EUAIAcExplainable}. This gap is particularly critical for generative AI systems deployed in high-stakes domains, where the primary risks arise from the underlying decision and operational processes, and where interface design critically shapes whether humans can effectively understand, monitor, and intervene in those processes, whose failures may have immediate and severe consequences. In many deployed AI systems, oversight is implemented as a single review step or a nominal approval interface, offering limited visibility into system behaviour and limited intervention authority \cite{Skitka1999AutomationBias, Mosier1997HighTechCockpits, Bucinca2021TrustOrThink, Green2022FlawsOversight}. Such designs can reduce intervention authority and visibility into system behaviour, and may foster over-reliance rather than effective control \cite{Bucinca2021TrustOrThink, Cummings2014ManVersusMachine, Bansal2021WholeExceedParts}. This gap between regulatory intent and technical implementation motivates our work.

We argue that effective human oversight requires architectural commitment across multiple system layers. Rather than treating HITL as a final validation gate, oversight mechanisms must be embedded within the system, automatically activated when risk signals exceed thresholds, and capable of supporting not only direct intervention (HITL) but also monitoring (HOTL) and governance through policy refinement \cite{HoDacMartinez2024HumanOversight, Sterz2024QuestEffectiveness, Manheim2018DynamicSafetyEnvelopes, Raji2020AccountabilityGap}. Effective oversight should ensure that decisions are logged with requirement identifiers, metric evidence, and rationales, and that human feedback is systematically used to refine system behaviour. 

This paper presents a preliminary, methodological contribution: a multi-layer oversight architecture for generative accessible UIs  that support high-stakes communication workflows. We instantiate the design in a healthcare communication setting,  with accessible post-consultation medication instructions tailored to user profiles, including individuals with cognitive disabilities and low vision, integrating deterministic rules, LLM-based generation, automated evaluation metrics, and structured human feedback into an end-to-end auditable pipeline. The proposed oversight mechanism is domain-agnostic \cite{Raji2020AccountabilityGap, Artsi2025LLMClinicalWorkflows}.  We do not report user study results; instead, we provide an implemented architecture that operationalizes meaningful human oversight in practice. We provide:

\begin{itemize}
\item An end-to-end oversight architecture that operationalizes escalation-driven HITL and HOTL across generation, evaluation, and governance.
\item A traceability scheme linking requirements, metric evidence, escalation decisions, and human rationales. 
\item A structured feedback loop that turns review outcomes into governance actions (rule and prompt updates, threshold calibration, and audit logs).
\end{itemize}

\section{Related work}

Human oversight has been studied across the human-automation interaction and HCI, commonly distinguishing direct intervention (Human-in-the-Loop, HITL) from supervisory control (Human-on-the-Loop, HOTL). Parasuraman et al. formalize levels of automation and show how misaligned allocation of control can harm performance and accountability \cite{Parasuraman2000Levels}. Human factors research also highlights automation bias and loss of situation awareness as recurring failure modes when humans supervise complex systems \cite{Parasuraman2000Levels, Endsley1995SituationAwareness, Schemmer2022ExplainableAIandAutomationBias, Echterhoff2024CognitiveBiasLLMs, LiAral2025HumanTrustAISearch}.

In HCI, Amershi et al. provide widely used guidelines for Human–AI Interaction, emphasizing timing, transparency, and user control \cite{Amershi2019GuidelinesHumanAI}. However, these guidelines primarily address interaction-level recommendations and leave open how to implement oversight as an enforceable mechanism across multi-stage generative pipelines \cite{Raji2020AccountabilityGap, Ferrari2025ObserveInspectModify}. Related work on supervisory control shows that intervention quality degrades under high workload and limited transparency \cite{Cummings2014ManVersusMachine}, and that explanation/control interfaces can increase over-reliance rather than calibrated trust \cite{Bucinca2021TrustOrThink, Bansal2021WholeExceedParts}. 

Recent syntheses of human–AI collaboration further highlight that many deployed systems still rely on limited interaction patterns and weak forms of collaborative control. Gomez et al. identify recurring structures of human–AI interaction in decision-support pipelines and show that collaboration often remains asymmetric and poorly integrated into workflow design \cite{gomez_human_ai_collabo_2025}. This reinforces the need for architectural approaches that explicitly structure intervention, supervision, and continuous improvements across system pipelines, rather than treating human involvement as a single interaction layer.

Accessibility standards and guidance (WCAG 2.2 \cite{W3C2023WCAG22}, EN 301 549 \cite{ETSI2021EN301549}, ISO 24495-1 \cite{ISO2023PlainLanguage}, W3C COGA \cite{W3C2021COGAUsable}) stress that information must be understandable, not only correct, including information presented to human reviewers and supervisors during oversight and escalation activities, aligning with Cognitive Load Theory’s account of how complexity and poor structure impair comprehension in safety-critical contexts \cite{Sweller1988CognitiveLoad, Lyell2018CognitiveLoadAutomationBias, Moreno2024DesigningUISimplification}.  Building on these foundations, our work operationalizes escalation-driven HITL and HOTL as an auditable architecture that links risk signals, intervention points, and structured feedback to ongoing governance \cite{Manheim2018DynamicSafetyEnvelopes, Raji2020AccountabilityGap}.  Automated metrics are frequently used to assess generated outputs. Semantic similarity metrics (e.g., BERTScore) \cite{Zhang2020BERTScore}, factual consistency measures (e.g., FactCC) \cite{Kryscinski2020FactualConsistencySummarization}, and readability indices can provide useful signals, but they are often insufficient proxies for comprehension, particularly for cognitively vulnerable users \cite{Sweller1988CognitiveLoad, Martinez2024ExploringLLMsEasyToRead, Alarcon2024EASIERSystem, Saeuberli2024DigitalComprehensibility}. We therefore treat metrics as risk detectors that trigger escalation rather than as final arbiters.

Concretely, this work extends our prior model-driven architecture for LLM-driven accessible interfaces \cite{Jerry2025LLMDrivenAccessibleInterface}, featuring SysML v2 traceability and standards-aligned templates, with escalation-driven HITL and HOTL oversight,  and governance mechanisms.

\section{System architecture overview}

We introduce a design-driven architecture for generating accessible user interfaces and UI content using large language models (LLMs). The approach is model-based and aligns with standards: it employs structured user profiles, declarative adaptation rules, and validated prompt templates to create multimodal accessible outputs (e.g., plain-language text, pictograms, high-contrast layouts) with clear traceability to normative requirements (e.g., WCAG, EN 301 549, ISO 24495-1, W3C COGA).

\begin{figure}
  \centering
  \includegraphics[width=\linewidth]{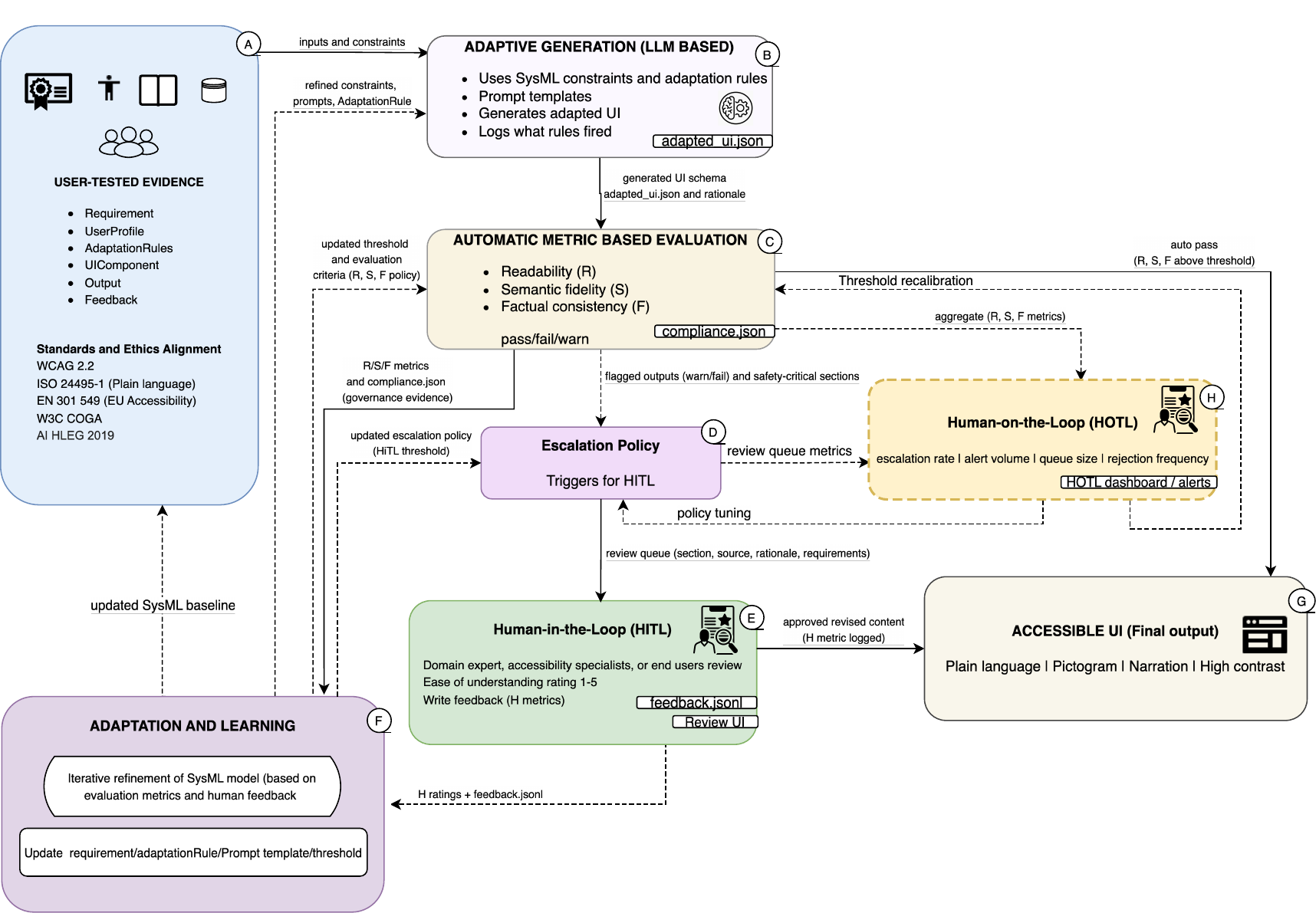}
  \caption{End-to-end architecture for accessible UI generation with layered human oversight}
  \label{fig:architecture}
\end{figure}
\autoref{fig:architecture} summarizes the end-to-end pipeline. (A) User-tested evidence captures user profiles, requirements, adaptation rules, and UI components, and anchors them to standards. This layer acts as the source of truth for what must be satisfied and what evidence must be logged. In prior work, SysML v2 is used to maintain explicit traceability between user needs, adaptation rules, and normative requirements, enabling auditable transformations in generative accessible interfaces. 

(B) Adaptive generation (LLM-based) produces a structured UI schema (e.g., headings, sections, modality choices, theming) guided by constraints derived from the requirement model and rule set. The goal is not free-form UI generation, but controlled generation bounded by (i) baseline accessible templates and (ii) profile-driven refinements (e.g., plain language, stepwise structuring, pictograms). 

(C) Automatic metric-based evaluation computes checkpoint signals for generated UI communication, including readability, semantic fidelity, and factual consistency. These signals are logged as machine-readable evidence (e.g., in a compliance.json artifact) and are used as risk indicators rather than final judgments. This is consistent with our POC pipeline that formalizes checkpoints R, S, F, H (R using Flesch-Szigriszt and Fernández–Huerta (Spanish) \cite{SzigrisztPazos1992Perspicuidad, FernandezHuerta1959Lecturabilidad}, S using BERTScore \cite{Zhang2020BERTScore}, F using AlignScore \cite{Zha2023AlignScore}, and H captured via Ease-of-Understanding (1–5) ratings with comments while logging evidence for auditability.

(D) Escalation policy translates risk signals into enforceable control: outputs that violate thresholds, contain safety-critical segments, or show high uncertainty are routed to mandatory human review prior to release. This is the architectural “commitment” that turns evaluation into governance, rather than leaving oversight as an afterthought.

(E) Human-in-the-Loop review is performed by the appropriate stakeholders (domain experts, accessibility specialists, and end users) using structured review instruments (e.g., ease-of-understanding ratings and component-level annotations). Feedback is recorded in a structured log (e.g., feedback.jsonl) linked to requirement IDs and UI nodes, enabling traceable corrective actions. The POC similarly captures human feedback as structured events and links it back to SysML requirement identifiers, closing the loop between human validation and model updates. 

Finally, (F) Adaptation and learning aggregates metric evidence and human feedback to update governance artifacts (rules, prompt templates, thresholds, and requirement refinements), while preserving versioned traceability.
We instantiate the architecture in healthcare communication (patient-facing interface content derived from clinical text), consistent with prior model-driven accessible UI generation work; however, the oversight mechanism (links risk signals, escalation, structured review, and governance updates) is domain-agnostic and can transfer to other high-stakes UI settings (e.g., legal or public services).

\section{Oversight mechanism}

Our oversight model separates intervention, supervision, and governance so responsibilities remain explicit and auditable.

\subsection{Oversight Modalities}

\begin{itemize}
\item \textbf {HITL (intervention).} HITL is activated by the escalation policy when automated signals are insufficient to guarantee safety or accessibility. Reviewers validate: (i) meaning preservation relative to the source, (ii) factual consistency for safety-relevant communication claims, and (iii) understandability for the target profile. Outputs include approval or rejection, targeted revision requests, and structured rationales linked to requirement and UI component identifiers.
In the healthcare communication use case, HITL reviewers act as clinical communication reviewers and accessibility reviewers for post-consultation medication instructions. Escalation is triggered, for example, when (a) readability or plain-language scores fall below thresholds for a cognitive-impairment profile, (b) contradictions are detected between generated instructions and structured medication data (e.g., dosage, timing, or route of administration), or (c) warnings related to known contraindications or safety advice are missing. Reviewers may edit or request revisions to the generated instruction text, confirm or correct safety-relevant statements, and validate that the presentation and wording satisfy accessibility and understandability requirements for the intended user profile.
\item \textbf {HOTL (supervision).} HOTL monitors the system’s behavior over time rather than reviewing every output. Supervisors inspect aggregate signals such as alert volume, escalation rates, failure modes by requirement category, and compliance evidence trends, and use these to recalibrate thresholds and detect drift.
In the healthcare communication setting, HOTL supervision focuses on escalation rates and failure patterns across user profiles (e.g., cognitive impairment or low-vision profiles) and across instruction types (e.g., single-drug versus multi-drug regimens). Drift is indicated, for example, by rising numbers of semantic mismatches between generated instructions and structured medication data, or by repeated reviewer overrides for the same classes of accessibility or safety requirements.
\item \textbf {Governance (continuous improvement).} Governance converts review outcomes into controlled updates to the system’s “policy surface”: adaptation rules, prompt templates, threshold settings, and requirement refinements. This mirrors the POC’s trace-first approach in which evaluation evidence and feedback are tied to requirement IDs to support auditable updates.
For the healthcare communication use case, governance actions include updating plain-language and accessibility adaptation rules, refining prompt templates for medication instruction generation, and recalibrating escalation thresholds based on recurring reviewer feedback for specific risk patterns (e.g., persistent readability violations for certain cognitive profiles or recurring safety-related omissions).
\end{itemize}
This healthcare communication use case is employed to ground the oversight architecture in a realistic, safety-sensitive communication scenario and does not imply that the system performs or replaces clinical decision-making.

\subsection{Escalation policy}

The escalation policy is explainable and auditable: it records which signals triggered review, which UI sections were flagged (including safety-relevant communication segments), and which stakeholder role is required for resolution. Human feedback is captured as structured, versioned events referencing the same identifiers used in the requirement model, enabling traceable governance actions and compliance reporting.

\section{Trade-offs and Roles}

Oversight must balance safety, scalability, and workload. The architecture avoids two extremes: (i) fully automated approval based on metrics alone, and (ii) universal manual review. Instead, it makes HITL selective and risk-proportional via escalation, while HOTL ensures continuous monitoring and calibration.
Stakeholder roles map to evidence types:
\begin{itemize}
\item End users: provide comprehension-oriented feedback (ease-of-understanding, usability notes) that standards alone cannot guarantee.
\item Domain experts: validate correctness and safety in escalated cases.
\item Accessibility specialists: verify standards-based constraints and profile fit.
\item System owners and governance: tune policy thresholds, review audit logs, and maintain traceability for compliance.
\end{itemize}

\section{Discussion and work in progress}

This work argues that meaningful oversight in generative IUIs is an architectural property: risk must be surfaced through UI-level signals, escalation must be enforceable, and decisions must be traceable to requirements and evidence. Our current contribution is methodological, demonstrating an implemented pipeline that integrates model-driven accessibility constraints with risk-based escalation and structured review. 

Limitations include (i) the need for empirical validation of how automated signals align with human judgments across profiles, and (ii) the need for longitudinal monitoring to study drift and policy recalibration in deployment. Next steps include user studies with target populations and domain professionals, and evaluating whether checkpoint thresholds and escalation strategies generalize across user-interface settings that support high-stakes workflows.

This workshop paper focuses on oversight-by-design and does not replicate our prior model-driven accessible-interface pipeline, which provides SysML v2 traceability and standards-aligned templates \cite{Jerry2025LLMDrivenAccessibleInterface}. We also draw on an ongoing proof-of-concept in healthcare communication (unpublished) as an instantiation context.

\section*{Acknowledgments}
This work has also been supported by grants PID2023-148577OB-C21
(Human-Centered AI: User Driven Adapted Language Models-HUMAN\_AI)
funded by MICIU/AEI/10.13039/501100011033
and by FEDER/UE.

\section*{Declaration on Generative AI}
 During the preparation of this work, the authors used ChatGPT and Grammarly in order to: Grammar and spelling check, paraphrase and reword. After using these tools, the authors reviewed and edited the content as needed and take full responsibility for the publication’s content.

\appendix

\end{document}